\documentclass[aps,prl,twocolumn,showpacs,amsmath,amssymb]{revtex4-1}
\usepackage{amsmath}
\usepackage{graphicx}
\usepackage{subfigure}
\usepackage{epstopdf}
\usepackage{color}
\usepackage{multirow}
\usepackage{setspace}
\usepackage{overpic}
\usepackage{amssymb}
\usepackage[bookmarksnumbered, pdfstartview=FitH,colorlinks,urlcolor=blue, citecolor=blue,linkcolor=blue] {hyperref}
\usepackage{lineno}
\usepackage{bm}
\usepackage{rotating}
\usepackage[utf8]{inputenc}

\let\oldequation\equation
\let\oldendequation\endequation

\renewenvironment{equation}
  {\linenomathNonumbers\oldequation}
  {\oldendequation\endlinenomath}

\begin{document}

\title{{\bf \boldmath Observation of $D^+_s\to \eta^\prime \mu^+\nu_\mu$, Precision Test of Lepton Flavor Universality with $D^+_s\to \eta^{(\prime)} \ell^+\nu_\ell$,
and First Measurements of $D^+_s\to \eta^{(\prime)}\mu^+\nu_\mu$ Decay Dynamics}}

\author{
M.~Ablikim$^{1}$, M.~N.~Achasov$^{13,b}$, P.~Adlarson$^{73}$, R.~Aliberti$^{34}$, A.~Amoroso$^{72A,72C}$, M.~R.~An$^{38}$, Q.~An$^{69,56}$, Y.~Bai$^{55}$, O.~Bakina$^{35}$, I.~Balossino$^{29A}$
% ...
\\
\vspace{0.2cm}
(BESIII Collaboration)\\
\vspace{0.2cm} {\it
$^{1}$ Institute of High Energy Physics, Beijing 100049, People's Republic of China\\
% ...
$^{13}$ G.I. Budker Institute of Nuclear Physics SB RAS (BINP), Novosibirsk 630090, Russia\\
% ...
$^{29}$ INFN Sezione di Ferrara, (A)INFN Sezione di Ferrara, I-44122, Ferrara, Italy; (B)University of Ferrara,  I-44122, Ferrara, Italy\\
% ...
$^{34}$ Johannes Gutenberg University of Mainz, Johann-Joachim-Becher-Weg 45, D-55099 Mainz, Germany\\
$^{35}$ Joint Institute for Nuclear Research, 141980 Dubna, Moscow region, Russia\\
% ...
$^{38}$ Liaoning Normal University, Dalian 116029, People's Republic of China\\
% ...
$^{55}$ Southeast University, Nanjing 211100, People's Republic of China\\
$^{56}$ State Key Laboratory of Particle Detection and Electronics, Beijing 100049, Hefei 230026, People's Republic of China\\
% ...
$^{69}$ University of Science and Technology of China, Hefei 230026, People's Republic of China\\
% ...
$^{72}$ University of Turin and INFN, (A)University of Turin, I-10125, Turin, Italy; (B)University of Eastern Piedmont, I-15121, Alessandria, Italy; (C)INFN, I-10125, Turin, Italy\\
$^{73}$ Uppsala University, Box 516, SE-75120 Uppsala, Sweden\\
% ...
$^{b}$ Also at the Novosibirsk State University, Novosibirsk, 630090, Russia\\
% ...
}
}


\begin{abstract}
By analyzing 7.33~fb$^{-1}$ of $e^+e^-$ annihilation data
collected at center-of-mass energies between 4.128 and 4.226~GeV with the BESIII detector,
we report the observation of the semileptonic decay $D^+_s\to
\eta^\prime \mu^+\nu_\mu$, with a statistical significance larger than 10$\sigma$, and the measurements of the $D_s^+ \to
\eta\mu^+\nu_\mu$ and $D_s^+ \to \eta^\prime\mu^+\nu_\mu$ decay dynamics for the first time.
The branching fractions of $D_s^+ \to \eta\mu^+\nu_\mu$ and $D_s^+ \to \eta^\prime\mu^+\nu_\mu$ are determined to be
$(2.235\pm0.051_{\rm stat}\pm0.052_{\rm syst})\%$ and
$(0.801\pm0.055_{\rm stat}\pm0.028_{\rm syst})\%$, respectively,
with precision improved by factors of 6.0 and 6.6 compared to the previous best measurements.
Combined with the results for the decays $D_s^+ \to \eta e^+\nu_e$ and $D_s^+ \to \eta^\prime e^+\nu_e$,
 the ratios of the decay widths are examined both inclusively and in several $\ell^+\nu_\ell$ four-momentum transfer ranges. No evidence for lepton flavor universality violation is found within the current statistics.
The products of the hadronic form factors $f_{+,0}^{\eta^{(\prime)}}(0)$ and the $c\to s$ Cabibbo-Kobayashi-Maskawa matrix element $|V_{cs}|$ are determined. The results based on the two-parameter series expansion are $f^{\eta}_{+,0}(0)|V_{cs}| = 0.452\pm0.010_{\rm stat}\pm0.007_{\rm syst}$ and $f^{\eta^{\prime}}_{+,0}(0)|V_{cs}| = 0.504\pm0.037_{\rm stat}\pm0.012_{\rm syst}$, which help to constrain present models on $f_{+,0}^{\eta^{(\prime)}}(0)$.
The forward-backward asymmetries are determined to be $\langle A_{\rm FB}^\eta\rangle=-0.059\pm0.031_{\rm stat}\pm0.005_{\rm syst}$ and $\langle A_{\rm FB}^{\eta^\prime}\rangle=-0.064\pm0.079_{\rm stat}\pm0.006_{\rm syst}$ for the first time, which are consistent with the theoretical calculation.
\end{abstract}

\maketitle

\oddsidemargin  -0.2cm
\evensidemargin -0.2cm

The couplings between the three families of leptons and the gauge bosons are expected to be equal in the standard model (SM). This property is known as lepton flavor universality (LFU). In recent years, however, hints of tensions between experimental measurements and the SM predictions were reported in the semileptonic (SL) $B$ decays~\cite{HFLAV}, the anomalous magnetic moment of the muon~\cite{Muong-2:2006rrc,Muong-2:2021ojo}, and the Cabibbo angle anomaly~\cite{Coutinho:2019aiy,Crivellin:2020lzu}.
For example, the measured branching fraction (BF) ratios
${\mathcal R}_{D^{(*)}}^{\tau/\ell}={\mathcal B}_{B\to \bar D^{(*)}\tau^+\nu_\tau}/{\mathcal B}_{B\to \bar D^{(*)}\ell^+\nu_\ell}$~($\ell=\mu$, $e$)~\cite{babar_1,babar_2,lhcb_1,belle2015,belle2016,Belle:2019rba,LHCb:2023zxo}
 deviate from the SM predictions by $3.3\sigma$~\cite{HFLAV}.
Although these tensions have been explained by various theoretical models~\cite{BhupalDev:2020zcy,Nomura:2021oeu,Bordone:2016gaq,Altmannshofer:2017yso,Crivellin:2017zlb,Becirevic:2016yqi,BFajfer2012,Fajfer2012,Celis2013,Crivellin2015,Crivellin2016,Bauer2016}, no definite conclusion is established yet. Precision tests of LFU in different SL decays of heavy mesons provide deeper insight into these anomalies.
Possible LFU in the SL $D^+_s$ decays is not yet well tested~\cite{Ke:2023qzc}, due to poor knowledge of the semimuonic $D^+_s$ decays. Reference~\cite{Fajfer2015} notes that there may indeed be observable LFU violation effects in the SL decays mediated via $c\to s\ell^+\nu_\ell$. In the SM, the ratio ${\mathcal R}^{\eta^{(\prime)}}_{\mu/e}={\mathcal B}_{D_s^+\to\eta^{(\prime)}\mu^+\nu_\mu}/{\mathcal B}_{D_s^+\to\eta^{(\prime)} e^+\nu_e}$ is predicted to be 0.95-0.99~\cite{Hu:2021zmy,Ivanov:2019nqd,Cheng:2017pcq}.
Precision measurements of $D^+_s\to \eta^{(\prime)}\mu^+\nu_\mu$ are important to test $\mu$-$e$ LFU with the SL decays $D^+_s\to \eta^{(\prime)}\ell^+\nu_\ell$. Especially, the $D^+_s\to \eta\ell^+\nu_\ell$ decay is expected to be the most competitive mode in the $D^+_s$ sector.

Furthermore, measurements of the $D^+_s\to \eta^{(\prime)}\ell^+\nu_\ell$ dynamics are important to determine the $c\to s$ Cabibbo-Kobayashi-Maskawa (CKM) matrix element $|V_{cs}|$ and the vector or scalar hadronic form factors (FFs) $f_{+,0}^{\eta^{(\prime)}}(q^2)$,  after incorporating the inputs from the SM global fit~\cite{PDG2022} or theoretical calculations~\cite{Hu:2021zmy,Colangelo:2001cv,Azizi:2010zj,Offen:2013nma,Duplancic:2015zna,Ivanov:2019nqd,Bali:2014pva,Verma:2011yw,Melikhov:2000yu,Soni:2018adu,Faustov:2019mqr}.
The obtained results are critical to test the unitarity of the CKM matrix and validate different theoretical calculations on FFs. To date, only the vector FFs of $f_{+}^{\eta^{(\prime)}}(q^2)$ has been measured by analyzing the $D^+_s\to \eta^{(\prime)}e^+\nu_e$ dynamics at BESIII~\cite{BESIII:2019qci,BESIII:2023ajr}. However, no $f_{0}^{\eta^{(\prime)}}(q^2)$ is available due to negligible lepton mass in $D^+_s\to \eta^{(\prime)}e^+\nu_e$.
The $D^+_s\to\eta^{(\prime)}\mu^+\nu_\mu$ decays offer unique test-bed to access $f_{0}^{\eta^{(\prime)}}(q^2)$ besides $|V_{cs}|$ and $f_{+}^{\eta^{(\prime)}}(q^2)$. Especially, Refs.~\cite{Faustov:2019mqr,Ivanov:2019nqd} state that
the forward-backward asymmetry parameters ($\langle A_{\rm FB}\rangle$), defined relative to helicity amplitudes, are partially dependent on the expected decay rates and hadronic FFs; and non-zero asymmetries in $D^+_s\to \eta^{(\prime)}\mu^+\nu_\mu$ are expected~\cite{Faustov:2019mqr,Ivanov:2019nqd}.
The obtained $f_{0}^{\eta^{(\prime)}}(q^2)$ and $\langle A_{\rm FB}\rangle$ are important to validate different theoretical calculations~\cite{Duplancic:2015zna,Faustov:2019mqr,Ivanov:2019nqd,Bali:2014pva,Soni:2018adu},
thereby improve the precision of lattice quantum chromodynamics (LQCD) calculations on the hadronic FFs of SL $D^+_s$ decays~\cite{Bali:2014pva},
which are currently suffering large uncertainties compared to their $D^{0(+)}$ counterparts~\cite{FermilabLattice:2022gku,Parrott:2022rgu,Chakraborty:2021qav,Na:2011mc,FermilabLattice:2004ncd}.
These will reversely help to determine the CKM matrix elements precisely, which are crucial to test the unitarity of the CKM matrix at higher precision~\cite{Koponen:2012di,Koponen:2013tua,Brambilla:2014jmp,Bailey:2012rr}.

Previously, only BESIII reported the BFs of $D^+_s\to\eta^{(\prime)}\mu^+\nu_\mu$~\cite{Ablikim:2018} with large uncertainties of $20\%~(51\%)$ using 0.482 fb$^{-1}$ of $e^+e^-$ collision data taken at a center-of-mass energy $E_{\rm CM}=4.009$ GeV.
Using 7.33~fb$^{-1}$ of $e^+e^-$ collision data taken at $E_{\rm CM}$ between 4.128 and 4.226 GeV with the BESIII
detector, we report the first observation of $D^+_s\to \eta^\prime \mu^+\nu_\mu$,
and the BFs of $D^+_s\to \eta^{(\prime)} \mu^+\nu_\mu$ are determined with  improved precision by about sixfold. Based on these, we test $\mu$-$e$ LFU in $D^+_s\to \eta^{(\prime)}\ell^+\nu_\ell$ decays in the full kinematic range and several $q^2$ intervals.
By analyzing the $D^+_s\to \eta^{(\prime)}\mu^+\nu_\mu$ dynamics, we determine $f_{0}^{\eta^{(\prime)}}(q^2)$
and $\langle A_{\rm FB}\rangle$ for the first time.
Charge-conjugate modes are implied throughout this Letter.

A description of the design and performance of the BESIII detector can be found in
Ref.~\cite{Ablikim2010345}.
About 83\% of the data analyzed in this Letter profits from an upgrade of the end cap time-of-flight system
with multi-gap resistive plate chambers with a time resolution of
60\,ps~\cite{Lxin,Gyingxiao}.
 Monte Carlo (MC) simulated events are generated with a
{\sc{geant4}}-based~\cite{Agostinelli:2002hh} simulation software, which includes
the geometric description~\cite{Huang:2022wuo} and a simulation of the response of the detector. An inclusive MC sample with an equivalent luminosity of 40 times that of the data is produced at $E_{\rm CM}$ between 4.128 and 4.226~GeV.
It includes open charm processes, initial state radiation (ISR) production of charmonium [$\psi(3770)$, $\psi(3686)$, and $J/\psi$], $q\bar q\,(q=u,\,d,\,s$) continuum processes, along with Bhabha
scattering, $\mu^+\mu^-$, $\tau^+\tau^-$, and $\gamma\gamma$ events.  The open charm processes are
generated using {\sc{conexc}}~\cite{Ping:2013jka}. The effects of ISR and final state radiation are included.
 Signal MC samples of the SL decays $D^+_s\to \eta^{(\prime)} \mu^+\nu_\mu$ are simulated with the two-parameter series expansion
model~\cite{Chikilev:1999zn}, with parameters obtained in this work.
 The input cross section of $e^+e^-\to D^\pm_sD^{*\mp}_s$ is taken from Ref.~\cite{crosssection}.
In the MC generation, known particle decays are generated by {\sc{evtgen}}~\cite{ref:evtgen} with the BFs taken from the Particle Data Group~\cite{PDG2022}, and other modes are generated using {\sc{lundcharm}} \cite{ref:lundcharm}.

In $e^+e^-$ collisions at $E_{\rm CM}$ between 4.128 and 4.226~GeV, the $D_s^+$ mesons are produced
predominantly via $e^+e^-\to D^\pm_sD_s^{*\mp}$~\cite{CLEO:2008ojp}.
Candidates in which one $D_s^-$ is fully reconstructed in one of the fourteen hadronic decay modes,
$D^-_s\to K^+K^-\pi^-$, $K^+K^-\pi^-\pi^0$, $K^0_SK^-$,
$K^0_SK^-\pi^0$,
$K^0_SK^0_S\pi^-$,
$K^0_SK^+\pi^-\pi^-$,
$K^0_SK^-\pi^+\pi^-$,
$\pi^+\pi^-\pi^-$,
$\eta_{\gamma\gamma}\pi^-$, $\eta_{\pi^0\pi^+\pi^-}\pi^-$,
$\eta^\prime_{\eta_{\gamma\gamma}\pi^+\pi^-}\pi^-$, $\eta^\prime_{\gamma\pi^+\pi^-}\pi^-$,
$\eta_{\gamma\gamma}\rho^-$, and
$\eta_{\pi^0\pi^+\pi^-}\rho^-$,
 are called the single-tag (ST) $D^-_s$.
 Those in which the ST $D^-_s$, the transition $\gamma(\pi^0)$ of the $D_s^{*\mp}$ decay, and the signal decays of $D^+_s\to\eta^{(\prime)}\mu^+\nu_\mu$
are simultaneously reconstructed are called double-tag~(DT) events.
Based on these, we determine the BF of the signal decay by
\begin{equation}
{\mathcal B}_{\rm sig} = N_{\rm DT}/(N^{\rm tot}_{\rm ST}\cdot \epsilon_{\gamma(\pi^0)\rm sig}),
\end{equation}
where $N^{\rm tot}_{\rm ST}=\sum_k N^k_{\rm ST}$ and $N_{\rm DT}$ are the total ST and DT yields in data summing over tag mode $k$,
and $\epsilon_{\gamma(\pi^0)\rm sig}$ is the effective signal efficiency of selecting $\gamma(\pi^0)\eta^{(\prime)}\mu^+\nu_\mu$
in the presence of ST $D^-_s$. The $\epsilon_{\gamma(\pi^0)\rm sig}$ is the averaged efficiency of $\epsilon_{\gamma\rm sig}$ and $\epsilon_{\pi^0\rm sig}$, and estimated by
$\sum_k\frac{N^k_{\rm ST}}{N^{\rm tot}_{\rm ST}}\frac{\epsilon^k_{\rm DT}}{\epsilon^k_{\rm ST}}$,
where $\epsilon^k_{\rm ST}$ and $\epsilon^k_{\rm DT}$ are the ST and DT efficiencies for the $k$-th tag mode, respectively.

For each tag mode, the ST yield is extracted from a fit to the corresponding invariant-mass spectrum of the ST candidates.
The selection criteria for all ST candidates are the same as Ref.~\cite{BESIII:2023ajr}, where detailed description can be found.
The total ST yield is $N^{\rm tot}_{\rm ST}=(817.0\pm3.4_{\rm stat})\times 10^3$.

In the presence of ST $D_s^-$, we select candidates for the transition $\gamma\,(\pi^0)$ from $D^{*+}_s$
decay and signal $D_s^+\to\eta^{(\prime)}\mu^+\nu_\mu$ among the unused particles recoiling against the ST $D_s^-$. 
In the signal decay, the $\eta$ is reconstructed via $\eta\to\gamma\gamma$ or $\eta\to\pi^0\pi^+\pi^-$ decay, and the $\eta^\prime$ is reconstructed via $\eta^\prime\to\eta_{\gamma\gamma}\pi^+\pi^-$ or $\gamma\pi^+\pi^-$ decay. 
Particle identification (PID) for pions combines the 
specific ionization information in the multilayer drift chamber and
the flight time in the time-of-flight system, and PID for muons further combine the energy deposited in the electromagnetic
calorimeter~(EMC).
 Likelihoods under various particle hypotheses (${\mathcal L}_i$, $i=e$, $\pi$, $\mu$, and $K$) are calculated.
Charged tracks satisfying ${\mathcal L}_\pi>0.001$, ${\mathcal L}_\pi>{\mathcal L}_K$ are assigned as pion candidates,
and satisfying
 ${\mathcal L}_\mu>0.001$, ${\mathcal L}_\mu>{\mathcal L}_e$, ${\mathcal L}_\mu>{\mathcal L}_K$,
and $E_{\rm EMC} \in (0.10, 0.28)$~GeV are assigned as muon candidates,
where $E_{\rm EMC}$ is the energy deposited in the EMC of muon candidates.
The selection criteria for the 
 transition $\gamma\,(\pi^0)$ and $\eta^{(\prime)}$ are the same as those in Ref.~\cite{BESIII:2023ajr}.
The energy and momentum of the missing neutrino of the signal SL decay are derived
as $E_{\nu_\mu} \equiv E_{\rm CM} - \Sigma_i E_i$ and $\vec{p}_{\nu_\mu} \equiv -\Sigma_i \vec{p}_{i}$, respectively, where
$E_i$ and $\vec{p}_i$ are the energy and momentum of the particle $i$, with $i$ running over
the ST $D^-_s$, transition $\gamma\,(\pi^0)$, $\eta^{(\prime)}$, and $\mu^+$.

The yield of  signal events is determined by a fit to the distribution of the kinematic variable
$ {\rm MM}^2 \equiv E_{\nu_\mu}^2/c^4 - |\vec{p}_{\nu_\mu}|^2/c^2$.
To improve the $\rm MM^2$ resolution, the candidate tracks, along with the missing neutrino, are subjected to a 3-constraint kinematic fit requiring energy and momentum conservation, constraining the invariant-mass of each  $D_s^\pm$ meson to the known $D_s^\pm$ mass~\cite{PDG2022}, and constraining the invariant-mass of the $D_s^-\gamma(\pi^0)$ or $D_s^+\gamma(\pi^0)$ combination to the known $D_s^{*\pm}$ mass~\cite{PDG2022}. The combination with the lowest $\chi^2$ is kept.
The $\chi^2$ for $D_s^+\to\eta^\prime_{\gamma\pi^+\pi^-}\mu^+\nu_\mu$ is required to be less than 30 to further suppress the non-$D_s^\pm D_s^{*\mp}$ backgrounds.

To suppress the backgrounds, the energy of any unused shower ($E_{\rm \gamma~extra }^{\rm max}$) in an event is required to be less  than 0.2\,GeV.
The DT candidates are vetoed if they contain any additional charged tracks ($N_{\rm extra}^{\rm char}$)  or $\pi^0$ reconstructed by two unused photons ($N_{\rm extra}^{\pi^0}$).
To further reject the peaking backgrounds from $D_s^+\to\eta^{(\prime)} \pi^+$ and $D_s^+\to\eta^{(\prime)} \pi^+\pi^0$,
the invariant-masses of $\eta^{(\prime)} \mu^+$, $M_{\eta^{(\prime)} \mu^+}$, are required to be less than 1.8~GeV/$c^2$ for both $D_s^+\to\eta\mu^+\nu_\mu$ and $D_s^+\to\eta^\prime\mu^+\nu_\mu$, and the invariant-masses of $\eta^{(\prime)} \nu_\mu$, $M_{\eta^{(\prime)} \nu_\mu}$, are required to be greater than 0.97~GeV/$c^2$ and 1.27~GeV/$c^2$ for $D_s^+\to\eta\mu^+\nu_\mu$ and $D_s^+\to\eta^\prime\mu^+\nu_\mu$, respectively.

\begin{figure}[htp]
\centering
\includegraphics[width=0.475\textwidth]{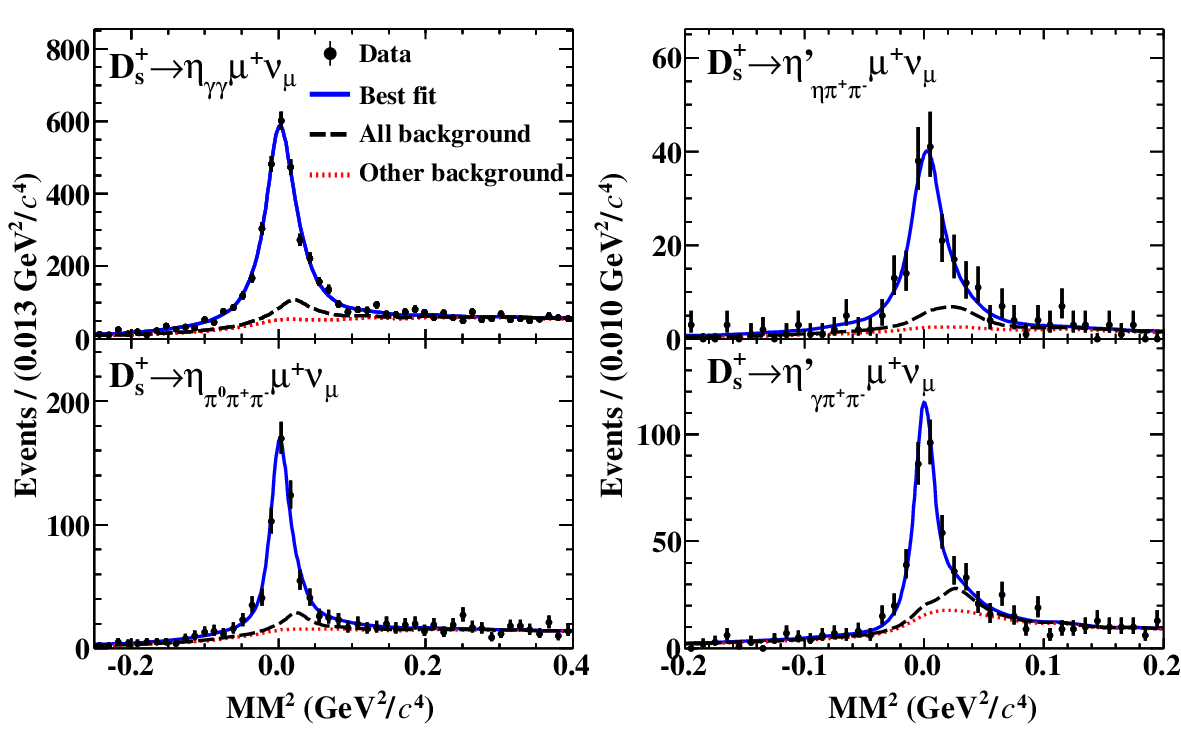}
\caption{Simultaneous fits to the MM$^2$ distributions of the accepted candidates for $D^+_s\to \eta^{(\prime)}\mu^+\nu_\mu$.
Points with error bars are data, solid curves are the best fits, differences between dashed and dotted curves are the $D_s^+\to\eta^{(\prime)} \pi^+(\pi^0)$ peaking backgrounds, and dotted curves are the other backgrounds.
\label{fig:signal_yields_fromdata}}
\end{figure}

After imposing all above selection criteria, the resulting MM$^2$ distributions of the accepted candidates for
different signal decay modes are exhibited in Fig.~\ref{fig:signal_yields_fromdata}.
For $D_s^+\to\eta^{(\prime)} \mu^+\nu_\mu$, the signal yield is extracted from a simultaneous unbinned maximum-likelihood fit
to the MM$^2$ spectra for the two $\eta$ or $\eta^\prime$ reconstruction modes,
with BFs constrained to be the same in the fit.
The signal, peaking backgrounds of $D_s^+\to\eta^{(\prime)}\pi^+(\pi^0)$, and other background shapes are modeled by the individual simulated shapes taken from the inclusive MC sample.  The signal and peaking background shapes are convolved with a Gaussian resolution function to account for the differences between data and simulation.  The yields of the peaking backgrounds are fixed to the expectation from simulation, and the other yields are left free.
The obtained signal efficiencies, signal yields, and resultant BFs are shown in Table~\ref{table:br}. The signal efficiencies have been corrected for small data-MC differences (overall factor $f^{\rm cor}=0.992 \sim 1.018$) in the $\eta, \pi^0$ reconstruction, $\mu^+$ PID, requirements of $E_{\gamma,~\rm extra}^{\rm max}, N_{\rm extra}^{\rm char}$, $N_{\rm extra}^{\pi^0}$,  $M_{\eta^{(\prime)}\mu^+}, M_{\eta^{(\prime)}\nu_\mu}$, and $\chi^2$.

   \begin{table}[hbtp]
   \centering
    \caption{\small Signal efficiencies ($\epsilon_{\rm \gamma(\pi^0){\rm sig}}$), signal yields ($N_{\rm DT}$), and obtained BFs ($\mathcal{B}_{\rm sig}$).  Efficiencies are averaged by $\epsilon_{\rm \gamma{\rm sig}}$ and $\epsilon_{\rm \pi^0{\rm sig}}$, and  include the BFs of the $\eta^{(\prime)}$ and $D_s^{*\mp}$ sub-decays.
    Numbers in the first and second parentheses are the most significant digits of the statistical and systematic uncertainties, respectively.  \label{table:br}}
           \begin{tabular}[t]{c|cc|cc}\hline\hline
Decay&\multicolumn{2}{c|}{$\eta\mu^+\nu_\mu$}&\multicolumn{2}{c}{$\eta^\prime\mu^+\nu_\mu$}\\
\hline
$\eta^{(\prime)}$ decay&$\gamma\gamma$& $\pi^0\pi^+\pi^-$&$\eta\pi^+\pi^-$&$\gamma\pi^+\pi^-$\\
$\epsilon_{\rm \gamma(\pi^0){\rm sig}}$~(\%)& 14.06(02)&2.89(01)&2.27(01)&3.64(01)\\
$N_{\rm DT}$&2567&528&149&238 \\
$\mathcal{B}_{\rm sig}$~(\%)&\multicolumn{2}{c|}{2.235(51)(52)}&\multicolumn{2}{c}{0.801(55)(28)}\\
\hline\hline
        \end{tabular}
\end{table}

The systematic uncertainties in the BF measurements are listed in Table 1 of Ref.~\cite{Supplement} and are discussed below.

The uncertainty in the ST $D^{-}_s$ yields is
studied by examining the change of the ST $D^{-}_s$ yields by varying the matched angle for signal shape and the order of Chebychev polynomial for background shape.
The uncertainties in the tracking or PID efficiencies of $\pi^\pm$ from the secondary $\eta^{(\prime)}$ decay and $\mu^+$ from primary $D_s^+$ decays are studied with the
control samples of $e^+e^-\to K^+K^-\pi^+\pi^-$ and $e^+e^-\to \gamma \mu^+\mu^-$, respectively.
The uncertainty of the $\pi^0$ or $\eta$ reconstruction is assigned
by studying the control sample of $e^+e^-\to K^+K^-\pi^+\pi^-\pi^0$.
The uncertainty in the reconstruction of transition $\gamma(\pi^0)$ from $D_s^{*+}$ is studied with the control sample of  $J/\psi\to \pi^0\pi^+\pi^-$~\cite{Ablikim:2011kv}.
The  uncertainty from the selection of the transition $\gamma(\pi^0)$ from $D_s^{*+}$ with the smallest $|\Delta E|$ method is estimated by using the control samples of $D_s^+\to K^+K^-\pi^+$ and $D_s^+\to\eta\pi^+\pi^0$.

The uncertainties due to the signal model are estimated
by comparing the DT efficiencies by varying the input hadronic FFs measured by $\pm 1\sigma$.
The uncertainties due to the $M_{\eta^{(\prime)}\mu^+}$ and $M_{\eta^{(\prime)}\nu_\mu}$ requirements are estimated by using the DT events of
$D_s^+\to\eta^{(\prime)}e^+\nu_e$.
The uncertainties from the $\eta^{(\prime)}$ reconstruction are estimated
by analyzing the control sample of $J/\psi\to \phi\eta^{(\prime)}$.
The systematic uncertainties in the ${\rm MM}^2$ fit are studied
by repeating the fits with different signal and background shapes.
The uncertainty in the peaking background yield is propagated
by varying its size by $\pm1 \sigma$ of the corresponding BF~\cite{PDG2022}.

The uncertainties due to different multiplicities of tag environments~\cite{Ablikim:2018jun} are assigned by studying of data-MC efficiency differences.
The uncertainty due to the finite MC statistics, which is dominated by that of the DT efficiency, is considered as a systematic uncertainty.

The uncertainties of the $E_{\rm \gamma~extra }^{\rm max}$, $N_{\rm extra}^{\pi^0}$, and
 $N_{\rm extra}^{\rm char}$ requirements are analyzed with  the
 DT events of $D_s^+\to\eta^{(\prime)}\pi^+(\pi^0)$ and $D_s^+\to\eta^{(\prime)}e^+\nu_e$.
The uncertainties of the $\chi^2$ requirements for $D_s^+\to\eta^{\prime}_{\gamma\pi^+\pi^-}\mu^+\nu_\mu$ are studied with
the DT events of $D_s^+\to\eta^{\prime}_{\gamma\pi^+\pi^-}\pi^+$ and $D_s^+\to\eta^{\prime}_{\gamma\pi^+\pi^-}e^+\nu_e$.
The uncertainties due to the quoted BFs of $\eta^{(\prime)}$ and $D_s^{*+}$ decays are taken from Ref.~\cite{PDG2022}.

The correlated and uncorrelated systematic uncertainties between the two $\eta^{(\prime)}$ decay modes are summarized in top and bottom section of Table 1 in Ref~\cite{Supplement}.
The combined systematic uncertainties are 2.3\% for $D_s^+\to\eta\mu^+\nu_\mu$ and 3.5\% for $D_s^+\to\eta^{\prime}\mu^+\nu_\mu$, taking into account  correlated and uncorrelated systematic uncertainties with the method described in Ref.~\cite{Schmelling:1994pz}.

The decay dynamics of $D^+_s\to\eta^{(\prime)} \mu^+\nu_\mu$ are investigated by dividing individual candidate events into $m=8(3)$ intervals of $q^2$ and performing a least-$\chi^2_{\rm FF}$ fit to the measured ($\Delta\Gamma^i_{\rm msr}$) and theoretically expected ($\Delta\Gamma^i_{\rm th}$) partial decay rates  in the $i$-th $q^2$ interval. The $\Delta\Gamma^i_{\rm th} \equiv\int_i\frac{d\Gamma}{dq^2}dq^2 $ relate to the hadronic FF via~\cite{Faustov:2019mqr}

\begin{widetext}
\begin{equation}
\begin{array}{l}
        \displaystyle \frac{d\Gamma}{dq^2} =
   \frac{G_{F}^{2}|V_{cs}|^{2}}{24\pi^{3}}\frac{(q^{2}-m^{2}_{\mu})^2|p_{\eta^{(\prime)}}|}{q^{4}m^{2}_{D_s^+}}
\left [(1+\frac{m^{2}_{\mu}}{2q^{2}})m^{2}_{D_s^+}|p_{\eta^{(\prime)}}|^2|f^{\eta^{(\prime)}}_{+}(q^{2})|^{2}
+\frac{3m^{2}_{\mu}}{8q^{2}}(m^{2}_{D_s^+}-m^{2}_{\eta^{(\prime)}})^{2}|f^{\eta^{(\prime)}}_{0}(q^{2})|^{2}\right ],
\end{array}
\end{equation}
\end{widetext}
where  $G_F$ is the Fermi coupling constant~\cite{PDG2022}, $|p_{\eta^{(\prime)}}|$ is the momentum of $\eta^{(\prime)}$ in the $D_s^+$ rest frame, $m_{\mu(\eta^{(\prime)})}$ is the $\mu^+(\eta^{(\prime)})$ mass. 
The hadronic FFs $f_+^{\eta^{(\prime)}}(q^2)$ are parameterized with a two-parameter series expansion~\cite{Becher:2005bg}.
We fix the pole mass at the known $D^{*+}_s$ mass~\cite{PDG2022}.
The similar formulas are applied for $f_0^{\eta^{(\prime)}}(q^2)$  
but with one-parameter series expansion due to much less contribution, and with the pole mass replaced with 
$m_{D_{s0}^{*}(2317)^+}$~\cite{PDG2022}.

The $\Delta\Gamma^i_{\rm msr}$ are determined by
$\Delta \Gamma^i_{\rm msr} = \frac{N_{\rm prd}^i}{\tau_{D_s^+}
  \cdot N^{\rm tot}_{\rm ST}}$, where
$\tau_{D_s^+}$ is the $D_s^+$ meson lifetime~\cite{PDG2022,Aaij:2017vqj}
and $N^i_{\rm prd}= \sum^{m}_{j}(\epsilon^{-1})_{ij} N_{\rm DT}^{j}$ is the corresponding produced signal yield.
 The observed signal yield ($N^j_{\rm DT}$) is obtained from a similar fit of the corresponding MM$^2$ distribution. The signal efficiency matrix ($\epsilon_{ij}$) is determined via
$\epsilon_{ij} = \sum_k\left[(1/N_{\rm ST}^{\rm tot})\cdot (N^{ij}_{\rm DT}/N^j_{\rm gen})_k
\cdot (N^k_{\rm ST}/\epsilon^k_{\rm ST})\cdot f^{\rm cor}\right]$,
where
$N^j_{\rm gen}$ is the total signal yield produced in the $j$-th $q^2$ interval,
$N^{ij}_{\rm rec}$ is the number of events generated in the $j\text{-}$th $q^{2}$ interval but reconstructed in the $i$-th $q^{2}$ interval,
and $k$ sums over all tag modes.
Details of $q^2$ divisions, the weighted signal efficiency matrices, $N^i_{\rm DT}$, $N^i_{\rm prd}$, and $\Delta\Gamma^i_{\rm msr}$ of different $q^2$ intervals  for $D^+_s\to \eta \mu^+\nu_\mu$ and $D^+_s\to \eta^\prime \mu^+\nu_\mu$ are shown in Tables 2-5 of Ref.~\cite{Supplement}, respectively.

The statistical and systematic covariance matrices are constructed as Ref.~\cite{BESIII:2023ajr}, which are shown in Tables 6-7 of Ref.~\cite{Supplement}.
The systematic covariance matrices include with those uncertainties from the BF measurements, along with the $D_s^+$ lifetime~\cite{PDG2022,Aaij:2017vqj}.

\begin{figure}
\centering
  \includegraphics[width=0.475\textwidth]{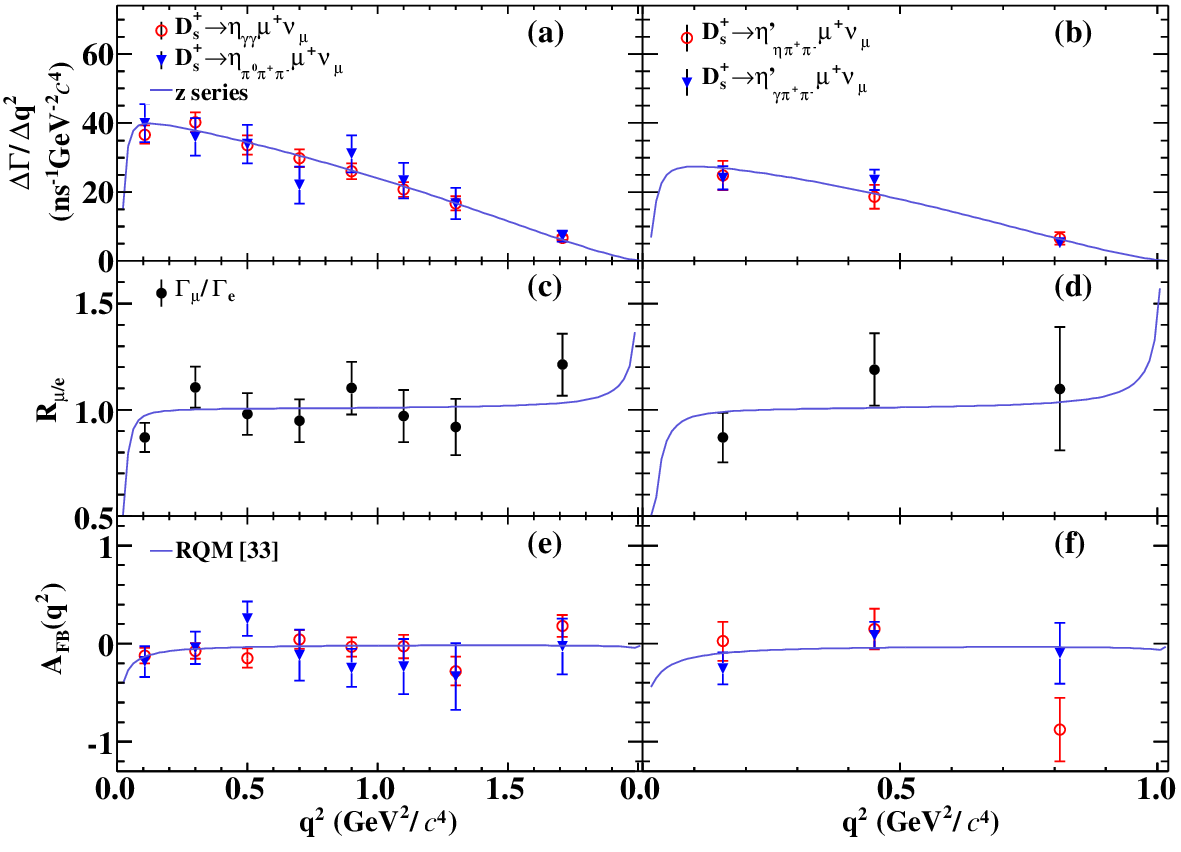}
\caption{\small
(a,b) Fits to $\Delta\Gamma^i_{\rm msr}$.
 (c,d) The measured $\mathcal R_{\mu/e}$ combining the two signal channels in each $q^2$ interval.
  (e,f) Comparisons of the measured $A_{\rm FB}$ and theoretical predications~\cite{Faustov:2019mqr}.
Red circles, blue triangles, and black points with error bars are data; the error bars combine both statistical and systematic
uncertainties. }
  \label{fig:combine_FF}
\end{figure}

\begin{figure}
\centering
  \includegraphics[width=0.475\textwidth]{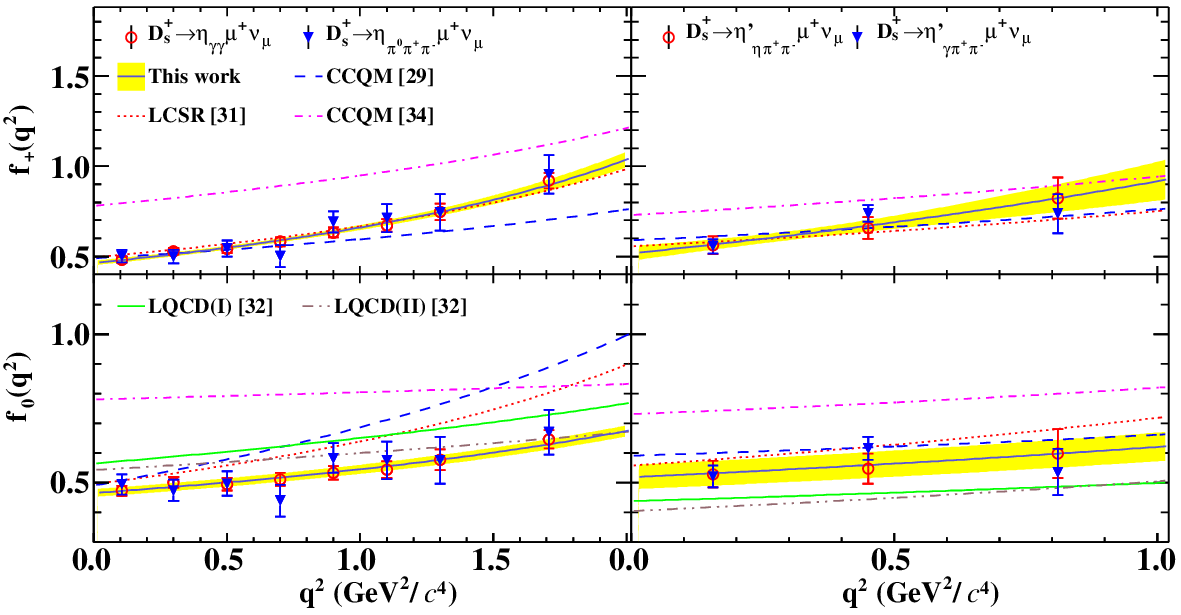}
\caption{\small
 Projections of the fits on $f_{+,0}^{\eta^{(\prime)}}(q^2)$. 
Red circles and blue triangles with error bars are data,
the yellow bands are the $\pm 1\sigma$ limits of fitted parameters, and the curves in different colors are from various theoretical calculations described in the legend.
}
  \label{fig:combine_fp}
\end{figure}

For each signal decay, we perform a simultaneous fit on the differential decay rates measured by the two $\eta^{(\prime)}$ sub-decays,
where the two modes are constrained to have same parameters for the hadronic FF.
Figures~\ref{fig:combine_FF}(a,b) show the fits to the differential decay rates.
The obtained parameters of hadronic FFs are summarized in Table~\ref{tab:FF_combine}.
Figure~\ref{fig:combine_fp} shows the projections to the extracted  $f_{+,0}^{\eta^{(\prime)}}(q^2)$, as well as the  comparison of $f_{+,0}^{\eta^{(\prime)}}(q^2)$ and various theoretical calculations.
Taking $|V_{cs}|$ from the SM global fit~\cite{PDG2022} as input,
we determine $f^{\eta}_{+,0}(0) = 0.465\pm0.010_{\rm stat}\pm0.007_{\rm syst}$ and $f^{\eta^{\prime}}_{+,0}(0)=0.518\pm0.038_{\rm stat}\pm0.012_{\rm syst}$.
Conversely, by taking the
$f^{\eta^{(\prime)}}_+(0)$ predicted by theory~\cite{Duplancic:2015zna} as inputs, we obtain $|V_{cs}|_\eta=0.913\pm0.020_{\rm stat}\pm0.014_{\rm syst}{^{+0.055}_{-0.053}}_{\rm theo}$ and $|V_{cs}|_{\eta^\prime}=0.904\pm0.067_{\rm stat}\pm0.021_{\rm syst}{^{+0.076}_{-0.073}}_{\rm theo}$, where the third uncertainties originate from the input FFs.
The obtained $f_+^{\eta^{(\prime)}}(0)$ are consistent with the relativistic quark model, QCD light-cone, and QCD sum rule calculations~\cite{Hu:2021zmy,Colangelo:2001cv,Azizi:2010zj,Offen:2013nma,Duplancic:2015zna,Faustov:2019mqr}. They disfavor the lattice QCD, covariant light-cone, and covariant confined quark model calculations~\cite{Ivanov:2019nqd,Bali:2014pva,Verma:2011yw,Melikhov:2000yu,Soni:2018adu} by more than $4\sigma$.

Combining the BFs measured in this work with our measurements
${\mathcal B}_{D^+_s\to \eta e^+\nu_e}=(2.255\pm0.039_{\rm stat}\pm0.051_{\rm syst})\%$ and
${\mathcal B}_{D^+_s\to \eta^\prime e^+\nu_e}=(0.810\pm0.038_{\rm stat}\pm0.024_{\rm syst})\%$~\cite{BESIII:2023ajr},
we obtain
${\mathcal R}^{\eta}_{\mu/e}=0.991\pm0.029_{\rm stat}\pm0.016_{\rm syst}$ and
${\mathcal R}^{\eta^\prime}_{\mu/e}=0.988\pm0.082_{\rm stat}\pm0.031_{\rm syst}$,
which are consistent with the SM predictions~\cite{Hu:2021zmy,Ivanov:2019nqd,Cheng:2017pcq}.
In addition, we examine the ${\mathcal R}^{\eta}_{\mu/e}$ and ${\mathcal R}^{\eta^\prime}_{\mu/e}$
in different $q^2$ intervals after considering the correlated uncertainties, with results shown in Figs.~\ref{fig:combine_FF}(c,d); these are also consistent with the SM predictions.

\begin{table}
\centering
\caption{\small Fitted parameters of hadronic FFs.
Quantities in the first (second) parentheses are the least two significant digits of statistical (systematic) uncertainties. NDF is the number of degrees of freedom. \label{tab:FF_combine}}
\begin{tabular}{lccc}\hline\hline
Decay  & $f_{+,0}^{\eta^{(\prime)}}(0)|V_{cs}|$ &  $r_1$ &$\chi^2_{\rm FF}$/NDF\\
\hline
$D_s^+\to\eta \mu^+ \nu_\mu$ &$0.452(10)(07)$  &$-2.9(06)(02)$ &5.1/14\\

$D_s^+\to\eta^\prime\mu^+\nu_\mu$& $0.504(37)(12)$   &$ -10.8(53)(14)$ &2.3/4 \\
                         \hline\hline
    \end{tabular}
\end{table}

The forward-backward asymmetry parameter $A_{\rm FB}$ is defined as $A_{\rm FB}(q^2)=\frac{\int_0^1d\cos\theta_\ell d\Gamma/d\cos\theta_\ell-\int_{-1}^0d\cos\theta_\ell d\Gamma/d\cos\theta_\ell}{\int_0^1d\cos\theta_\ell d\Gamma/d\cos\theta_\ell+\int_{-1}^0d\cos\theta_\ell d\Gamma/d\cos\theta_\ell}$, where $\theta_\ell$ is the angle between the momentum of the $\mu^+$ in the rest frame of the  $W$-boson and the
direction to the $W$-boson momentum in the rest frame of $D_s^+$.
The measured $A_{\rm FB}$ in various $q^2$  intervals are shown in Figs.~\ref{fig:combine_FF}(e,f).
The averaged $\langle A_{\rm FB}\rangle$ are determined to be $-0.059\pm0.031_{\rm stat}\pm0.005_{\rm syst}$ for $D_s^+\to\eta\mu^+\nu_\mu$ and $-0.064\pm0.079_{\rm stat}\pm0.006_{\rm syst}$ for $D_s^+\to\eta^\prime\mu^+\nu_\mu$, which are consistent with theoretical predictions~\cite{Faustov:2019mqr,Ivanov:2019nqd}.

In summary, we present for the first time the observation of $D^+_s\to \eta^\prime \mu^+\nu_\mu$ with statistical significance greater than 10$\sigma$ and the studies of $D^+_s\to \eta^{(\prime)}\mu^+\nu_\mu$ dynamics. The BFs of $D^+_s\to \eta \mu^+\nu_\mu$ and $D^+_s\to \eta^\prime \mu^+\nu_\mu$ are measured with precision improved by about sixfold over the previous best measurements~\cite{PDG2022}.  Combining with the BESIII measurements of $D_s^+\to\eta^{(\prime)}e^+\nu_e$~\cite{BESIII:2023ajr}, we calculate the $\mathcal R_{\mu/e}^{\eta^{(\prime)}}$ ratios in separate $q^2$ intervals and in the full range.
No significant evidence of LFU violation is found with current statistics.

By analyzing the dynamics of these decays, we determine
the hadronic FFs of $f_{+,0}^{\eta^{(\prime)}}(0)$, $|V_{cs}|$, the shapes of $f^{\eta^{(\prime)}}_{+,0}(q^2)$,
and  forward-backward asymmetry parameters of  $\langle A_{\rm FB}\rangle$.
The obtained $f^{\eta^{(\prime)}}_{0}(q^2)$ lineshapes offer crucial data to calibrate the $q^2$ dependent FFs from different theories for the first time.
 Unlike the comparable FFs in the SL decays mediated via $c\to d e^+\nu_e$~\cite{BESIII:2018xre},  the $D^+_s\to \eta$ FF measured in this work deviate from $f_+^{D\to K}(0)=0.7327\pm0.0049$ obtained via $D^0\to K^-\mu^+\nu_\mu$ ~\cite{BESIII:2018ccy} by more than $5\sigma$.
This rules out the expectation for comparable FFs for $D^{0(+)}\to \bar K \mu^+\nu_\mu$ and $D^+_s\to \eta \mu^+\nu_\mu$~\cite{Koponen:2012di},
but supports  that the spectator quarks play important role in FFs due to effects of confinement in the considered weak transitions and the SU(4) asymmetry breaking~\cite{Khlopov:1978id}. 

\begin{acknowledgments}

The authors thank Prof. Haibing Fu, Prof. Shan Cheng, Prof. Xianwei Kang, and Prof. Khlopov for helpful dis- cussions.
The BESIII Collaboration thanks the staff of BEPCII and the IHEP computing center for their strong support. This work is supported in part by National Key R\&D Program of China under Contracts No. 2020YFA0406400, No. 2023YFA1606000, No. 2023YFA1606704, and No. 2020YFA0406300; National Natural Science Foundation of China (NSFC) under Contracts No. 12305089, No. 12375092, No. 11635010, No. 11735014, No. 11835012, No. 11935015, No. 11935016, No. 11935018, No. 11961141012, No. 12022510, No. 12025502, No. 12035009, No. 12035013, No. 12061131003, No. 12192260, No. 12192261, No. 12192262, No. 12192263, No. 12192264, No. 12192265; 
the Chinese Academy of Sciences (CAS) Large-Scale Scientific Facility Program; the CAS Center for Excellence in Particle Physics (CCEPP); Joint Large-Scale Scientific Facility Funds of the NSFC and CAS under Contract No. U1932102, U1832207; CAS Key Research Program of Frontier Sciences under Contracts Nos. QYZDJ-SSW-SLH003, QYZDJ-SSW-SLH040; 100 Talents Program of CAS; 
Jiangsu Funding Program for Excellent Postdoctoral Talent under Contracts No. 2023ZB833; 
Project funded by China Postdoctoral Science Foundation under Contracts No. 2023M732547;
The Institute of Nuclear and Particle Physics (INPAC) and Shanghai Key Laboratory for Particle Physics and Cosmology; ERC under Contract No. 758462; European Union's Horizon 2020 research and innovation programme under Marie Sklodowska-Curie grant agreement under Contract No. 894790; German Research Foundation DFG under Contracts Nos. 443159800, 455635585, Collaborative Research Center CRC 1044, FOR5327, GRK 2149; Istituto Nazionale di Fisica Nucleare, Italy; Ministry of Development of Turkey under Contract No. DPT2006K-120470; National Research Foundation of Korea under Contract No. NRF-2022R1A2C1092335; National Science and Technology fund; National Science Research and Innovation Fund (NSRF) via the Program Management Unit for Human Resources \& Institutional Development, Research and Innovation under Contract No. B16F640076; Polish National Science Centre under Contract No. 2019/35/O/ST2/02907; Suranaree University of Technology (SUT), Thailand Science Research and Innovation (TSRI), and National Science Research and Innovation Fund (NSRF) under Contract No. 160355; The Royal Society, UK under Contract No. DH160214; The Swedish Research Council; U. S. Department of Energy under Contract No. DE-FG02-05ER41374.

%

\end{acknowledgments}

\clearpage
\appendix
\twocolumngrid
\setcounter{table}{0}
\setcounter{figure}{0}

\section*{Appendices}\label{Supplement}

Table~\ref{sys} summarizes the sources of the systematic uncertainties in the measurements of the branching fractions of $D_s^+\to\eta^{(\prime)}\mu^+\nu_\mu$. In this table, the contributions to the systematic uncertainties listed in the top part are treated as correlated, while those in the bottom part are treated as uncorrelated.

Tables~\ref{tab:effmatrixa} and~\ref{tab:effmatrixb} give the weighted efficiency matrices averaged over all fourteen ST modes for $D^+_s\to \eta \mu^+\nu_\mu$ and $D^+_s\to \eta^\prime \mu^+\nu_\mu$, respectively.

Tables~\ref{tab:decayratea} and ~\ref{tab:decayrateb} present the numbers of the  reconstructed events $N_{\rm DT}^i$ in data obtained from the $\rm MM^2$ fits, the numbers of the  produced events $N_{\rm prd}^i$, and the measured partial decay rate $\Delta\Gamma_{\rm msr}^i$  in different $q^2$ intervals for $D^+_s\to \eta \mu^+\nu_\mu$ and $D^+_s\to \eta^\prime \mu^+\nu_\mu$, respectively.

Table~\ref{tab:cova} summarizes the statistical correlation matrices and relative uncertainties for the measured partial decay rates in different $q^2$ intervals for $D^+_s\to \eta \mu^+\nu_\mu$. 

Table ~\ref{tab:covb} summarizes the systematic correlation matrices and relative uncertainties for the measured partial decay
rates in different $q^2$ intervals for $D^+_s\to \eta^\prime \mu^+\nu_\mu$.

\begin{table*}[htp]
\centering
\caption{Relative systematic uncertainties (in \%) on the measurements of the branching fractions
of $D_s^+\to \eta \mu^+\nu_\mu$ and $D_s^+\to \eta^\prime \mu^+\nu_\mu$. The top and the bottom sections are correlated and uncorrelated, respectively. The uncertainty in the uncorrelated $\pi^\pm$ tracking is obtained as the square root of the quadratic difference of the total uncertainty in the $\pi^\pm$ tracking and the correlated portion. The last row of combined uncertainties  are total uncertainties of $D_s^+\to \eta \mu^+\nu_\mu$ and $D_s^+\to \eta^\prime \mu^+\nu_\mu$ after taking into account correlated and uncorrelated systematic uncertainties.
}
\begin{tabular}{lcc|cc}
  \hline
  \hline
  Source  & $\eta_{\gamma\gamma}\mu^+\nu_\mu$&$\eta_{\pi^0\pi^+\pi^-}\mu^+\nu_\mu$&$\eta^\prime_{\eta\pi^+\pi^-}\mu^+\nu_\mu$&$\eta^\prime_{\gamma\pi^+\pi^-}\mu^+\nu_\mu$  \\
  \hline
ST $D^{-}_s$ yields                                      &0.5  &0.5  &0.5 &0.5    \\
  $\mu^+$ tracking                                             &0.3 &0.3 & 0.3&0.3        \\
  $\mu^+$ PID                                                  &0.3 &0.3 & 0.3&0.3        \\
     $\pi^\pm$ tracking                                           &--&1.2&0.6&0.6\\
  $\pi^\pm$ PID                                                &--&0.4&0.4&0.4\\

    $\pi^0$ or $\eta$ reconstruction                                  &1.1  &1.1  &0.8 &--    \\
 Transition $\gamma(\pi^0)$ reconstruction      &1.0&1.0&1.0&1.0\\
 Smallest $|\Delta E|$                                            &1.0&1.0&1.0&1.0\\
  Signal model                                         &0.3&0.3&0.6 &0.6        \\
    $M_{\rm \eta^{(\prime)}\mu^+}$ and $M_{\rm \eta^{(\prime)}\nu_\mu}$ requirements                                 &0.7&0.7&2.4&2.4         \\
Peaking background &0.7&0.7&1.0&1.0\\

\hline
   $\pi^\pm$ tracking                                           &--&--&1.7&--\\
$\eta^{(\prime)}$ reconstruction &--&0.1&0.1&1.4\\
  $\rm MM^2$ fit                                                    & 0.2&0.7&1.2&1.2          \\

  Tag bias                                                     &0.5&0.2 &0.2&0.2         \\
    MC  statistics                                               &0.3     &0.3&0.3&0.3           \\
   $E_{\rm \gamma~extra }^{\rm max}$,  $N_{\rm extra}^{\rm char}$, and $N_{\rm extra}^{\pi^0}$ requirements &0.4&0.8&0.8&1.2      \\

  $\chi^2$ requirement&--&--&--&1.6\\

  Quoted branching fractions                                   &0.5&1.1&1.3&1.4\\
  \hline
  Total                                                        &2.3           &3.0  &4.2&    4.4    \\
  \hline
  Combined&\multicolumn{2}{c|}{2.3}&\multicolumn{2}{c}{3.5}\\

  \hline
  \hline
\end{tabular}
\label{sys}
\end{table*}

\begin{table*}[htbp]\centering
\caption{The efficiency matrices for $D_s^+\to\eta\mu^+\nu_\mu$ averaged over all fourteen ST modes, where $\varepsilon_{ij}$ represents the
efficiency of events produced in the $j$-th $q^2$ interval and reconstructed in the $i$-th $q^2$ interval.}
\label{tab:effmatrixa}
\begin{tabular}{c|cccccccc|cccccccc}\hline\hline
\multirow{2}{*}{$\varepsilon_{ij}~(\%)$}&\multicolumn{8}{c|}{$D_s^+\to\eta_{\gamma\gamma}\mu^+\nu_\mu$}&\multicolumn{8}{c}{$D_s^+\to\eta_{\pi^0\pi^+\pi^-}\mu^+\nu_\mu$}\\
&1&2&3&4&5&6&7&8&1&2&3&4&5&6&7&8\\
\hline
1&13.18&0.80&0.02&0.00&0.00&0.00&0.00&0.00&2.99&0.14&0.00&0.00&0.00&0.00&0.00&0.00\\
2&1.05&12.10&1.01&0.03&0.00&0.00&0.00&0.00&0.21&2.70&0.17&0.01&0.00&0.00&0.00&0.00\\
3&0.04&1.16&11.87&1.12&0.04&0.00&0.00&0.00&0.01&0.23&2.60&0.17&0.01&0.00&0.00&0.00\\
4&0.01&0.05&1.27&11.84&1.13&0.04&0.00&0.00&0.00&0.01&0.27&2.53&0.17&0.01&0.00&0.00\\
5&0.00&0.01&0.05&1.32&11.64&1.11&0.02&0.00&0.00&0.00&0.02&0.27&2.38&0.15&0.01&0.00\\
6&0.00&0.00&0.01&0.06&1.22&11.45&1.09&0.01&0.00&0.00&0.00&0.02&0.26&2.20&0.17&0.00\\
7&0.00&0.00&0.00&0.01&0.05&1.15&11.35&0.59&0.00&0.00&0.00&0.01&0.02&0.23&2.08&0.09\\
8&0.00&0.00&0.01&0.01&0.02&0.05&1.00&12.83&0.00&0.00&0.00&0.00&0.00&0.02&0.22&2.14\\
\hline\hline
\end{tabular}
\end{table*}

\begin{table*}[htbp]\centering
\caption{
The efficiency matrices for $D_s^+\to\eta^\prime\mu^+\nu_\mu$ averaged over all fourteen ST modes, where $\varepsilon_{ij}$ represents the
efficiency of events produced in the $j$-th $q^2$ interval and reconstructed in the $i$-th $q^2$ interval.
}
\label{tab:effmatrixb}
\begin{tabular}{c|ccc|ccc}\hline\hline
\multirow{2}{*}{$\varepsilon_{ij}~(\%)$}&\multicolumn{3}{c|}{$D_s^+\to\eta^\prime_{\eta\pi^+\pi^-}\mu^+\nu_\mu$}&\multicolumn{3}{c}{$D_s^+\to\eta^\prime_{\gamma\pi^+\pi^-}\mu^+\nu_\mu$}\\
&1&2&3&1&2&3\\
\hline
1&2.11&0.10&0.00&3.28&0.12&0.04\\
2&0.08&2.11&0.13&0.11&3.49&0.17\\
3&0.00&0.07&2.35&0.04&0.11&3.91\\
\hline\hline
\end{tabular}
\end{table*}

\begin{table*}[htp]\centering
\caption{The partial decay rates of $D_s^+\to\eta\mu^+\nu_\mu$ in different $q^{2}$ intervals of data, where the uncertainties of partial decay rates are statistical only. }
\label{tab:decayratea}
\scalebox{0.93}{
\begin{tabular}{cccccccccc}\hline\hline
\multicolumn{2}{c}{$i$}&1&2&3&4&5&6&7&8\\
\multicolumn{2}{c}{$q^2$ $(\mathrm{GeV}^{2}/c^{4})$}&($m_\mu^2,\,0.2$)&($0.2,\,0.4$)&($0.4,\,0.6$)&($0.6,\,0.8$)&($0.8,\,1.0$)&($1.0,\,1.2$)&($1.2,\,1.4$)&($1.4,\,q_{\rm max}^2$)\\
\hline
\multirow{3}{*}{$D_s^+\to\eta_{\gamma\gamma} \mu^+\nu_\mu$}&$N_{\mathrm{DT}}^i$&403(26)&459(28)&397(27)&352(24)&303(21)&239(19)&188(19)&235(20)\\
&$N_{\mathrm{prd}}^i$&2850(203)&3309(235)&2768(233)&2454(206)&2142(184)&1712(173)&1377(170)&1713(160)\\
&$\Delta\Gamma^i_{\rm msr}$ $(\mathrm{ns^{-1}})$&6.92(49)&8.04(57)&6.72(57)&5.96(50)&5.20(45)&4.16(42)&3.34(41)&4.16(39)\\
\hline
\multirow{3}{*}{$D_s^+\to\eta_{\pi^0\pi^+\pi^-} \mu^+\nu_\mu$}&$N_{\mathrm{DT}}^i$&97(13)&91(12)&83(12)&58(11)&70(10)&52(09)&35(08)&43(09)\\
&$N_{\mathrm{prd}}^i$&3108(428)&2966(449)&2790(465)&1818(442)&2565(441)&1921(422)&1370(376)&1843(414)\\
&$\Delta\Gamma^i_{\rm msr}$ $(\mathrm{ns^{-1}})$&7.55(104)&7.20(109)&6.78(113)&4.41(107)&6.23(107)&4.67(103)&3.33(091)&4.48(100)\\

\hline\hline
\end{tabular}
}
\end{table*}

\begin{table*}[htp]\centering
\caption{
The partial decay rates of $D_s^+\to\eta^\prime \mu^+\nu_\mu$ in different $q^{2}$ intervals. Numbers in the parentheses are the statistical uncertainties. }
\label{tab:decayrateb}

\scalebox{0.95}{
\begin{tabular}{ccccc}\hline\hline
\multicolumn{2}{c}{$i$}&1&2&3\\
\multicolumn{2}{c}{$q^2$ $(\mathrm{GeV}^{2}/c^{4})$}&($m_\mu^2,\,0.3$)&($0.3,\,0.6$)&($0.6,\,q_{\rm max}^2$)\\
\hline
\multirow{3}{*}{$D_s^+\to\eta^\prime_{\eta\pi^+\pi^-} \mu^+\nu_\mu$}&$N_{\mathrm{DT}}^i$&64(11)&53(09)&28(07)\\
&$N_{\mathrm{prd}}^i$&2944(503)&2300(425)&1138(317)\\
&$\Delta\Gamma^i_{\rm msr}$ $(\mathrm{ns^{-1}})$&7.15(122)&5.58(103)&2.76(077)\\

\hline
\multirow{3}{*}{$D_s^+\to\eta^\prime_{\gamma\pi^+\pi^-} \mu^+\nu_\mu$}&$N_{\mathrm{DT}}^i$&98(13)&106(13)&40(10)\\
&$N_{\mathrm{prd}}^i$&2866(402)&2906(365)&912(265)\\
&$\Delta\Gamma^i_{\rm msr}$ $(\mathrm{ns^{-1}})$&6.96(98)&7.06(89)&2.22(64)\\

\hline\hline
\end{tabular}
}

\end{table*}

\begin{table*}[htp]\centering
\caption{Statistical and systematic correlation matrices and relative uncertainties for the measured partial decay rates of $D_s^+\to\eta \mu^+\nu_\mu$ in different $q^2$ intervals.
}
\label{tab:cova}
\scalebox{0.9}{
\begin{tabular}{ccccccccccccccccc}\hline\hline
\multicolumn{17}{c}{Statistical correlation matrix}\\
\multirow{2}{*}{$\rho_{ij}^{\rm stat}$}&\multicolumn{8}{c}{$D_s^+\to\eta_{\gamma\gamma}\mu^+\nu_\mu$}&\multicolumn{8}{c}{$D_s^+\to\eta_{\pi^0\pi^+\pi^-}\mu^+\nu_\mu$}\\
&1&2&3&4&5&6&7&8&1&2&3&4&5&6&7&8\\
\hline
1	&1.000	&-0.144	&0.015	&-0.002	&0.000	&0.000	&0.000	&0.000	&0.000	&0.000	&0.000	&0.000&0.000	&0.000	&0.000	&0.000	\\
2	&&1.000	&-0.180	&0.022	&-0.003	&0.000	&0.000	&0.000	&0.000	&0.000	&0.000	&0.000	&0.000&0.000	&0.000	&0.000	\\
3	&&&1.000	&-0.203	&0.025	&-0.003	&0.000	&0.000	&0.000	&0.000	&0.000	&0.000	&0.000&0.000	&0.000	&0.000	\\
4	&&&&1.000	&-0.209	&0.023	&-0.003	&0.000	&0.000	&0.000	&0.000	&0.000	&0.000	&0.000&0.000	&0.000	\\
5	&&&&&1.000	&-0.200	&0.023	&-0.003	&0.000	&0.000	&0.000	&0.000	&0.000	&0.000	&0.000&0.000	\\
6	&&&&&&1.000	&-0.194	&0.016	&0.000	&0.000	&0.000	&0.000	&0.000	&0.000	&0.000	&0.000\\
7	&&&&&&&1.000	&-0.130	&0.000	&0.000	&0.000	&0.000	&0.000	&0.000	&0.000	&0.000	\\
8	&&&&&&&&1.000	&0.000	&0.000	&0.000	&0.000	&0.000	&0.000	&0.000	&0.000	\\
1	&&&&&&&&&1.000	&-0.120	&0.010	&-0.001	&0.000	&0.000	&0.000	&0.000	\\
2	&&&&&&&&&&1.000	&-0.150	&0.015	&-0.002	&0.000	&0.000	&-0.001	\\
3	&&&&&&&&&&&1.000	&-0.173	&0.013	&-0.003	&0.000	&-0.002	\\
4	&&&&&&&&&&&&1.000	&-0.179	&0.015	&-0.005	&0.000	\\
5	&&&&&&&&&&&&&1.000	&-0.182	&0.014	&-0.002	\\
6	&&&&&&&&&&&&&&1.000	&-0.188	&0.010	\\
7	&&&&&&&&&&&&&&&1.000	&-0.135	\\
8	&&&&&&&&&&&&&&&&1.000	\\

\hline
Stat (\%)&7.11&7.09&8.41&8.38&8.60&10.11&12.34&9.37&13.78&15.13&16.65&24.32&17.20&21.98&27.48&22.43\\

\hline
\multicolumn{17}{c}{Systematic correlation matrix}\\
\multirow{2}{*}{$\rho_{ij}^{\rm syst}$}&\multicolumn{8}{c}{$D_s^+\to\eta_{\gamma\gamma}\mu^+\nu_\mu$}&\multicolumn{8}{c}{$D_s^+\to\eta_{\pi^0\pi^+\pi^-}e^+\nu_e$}\\
&1&2&3&4&5&6&7&8&1&2&3&4&5&6&7&8\\
\hline
1	&1.000	&0.725	&0.748	&0.698	&0.528	&0.596	&0.664	&0.609	&0.499	&0.501	&0.442	&0.460&0.543	&0.550	&0.479	&0.511	\\
2	&&1.000	&0.748	&0.409	&0.748	&0.251	&0.660	&0.416	&0.654	&0.273	&0.692	&0.217	&0.300&0.528	&0.550	&0.535	\\
3	&&&1.000	&0.588	&0.646	&0.428	&0.635	&0.491	&0.554	&0.474	&0.671	&0.552	&0.430&0.541	&0.441	&0.438	\\
4	&&&&1.000	&0.223	&0.778	&0.521	&0.595	&0.227	&0.643	&0.190	&0.701	&0.624	&0.430&0.245	&0.296	\\
5	&&&&&1.000	&0.056	&0.576	&0.131	&0.605	&0.195	&0.687	&0.167	&0.200	&0.444	&0.460&0.436	\\
6	&&&&&&1.000	&0.373	&0.563	&0.103	&0.580	&-0.011	&0.587	&0.586	&0.335	&0.166	&0.220\\
7	&&&&&&&1.000	&0.392	&0.476	&0.368	&0.448	&0.316	&0.393	&0.454	&0.416	&0.419	\\
8	&&&&&&&&1.000	&0.267	&0.409	&0.192	&0.382	&0.429	&0.344	&0.261	&0.283	\\
1	&&&&&&&&&1.000	&0.481	&0.799	&0.253	&0.421	&0.654	&0.630	&0.729	\\
2	&&&&&&&&&&1.000	&0.373	&0.714	&0.698	&0.584	&0.349	&0.557	\\
3	&&&&&&&&&&&1.000	&0.311	&0.311	&0.598	&0.570	&0.596	\\
4	&&&&&&&&&&&&1.000	&0.657	&0.504	&0.266	&0.284	\\
5	&&&&&&&&&&&&&1.000	&0.572	&0.487	&0.519	\\
6	&&&&&&&&&&&&&&1.000	&0.540	&0.649	\\
7	&&&&&&&&&&&&&&&1.000	&0.568	\\
8	&&&&&&&&&&&&&&&&1.000	\\
\hline
Syst (\%)&3.07&3.35&3.55&3.35&3.71&3.50&2.98&3.56&3.70&4.20&5.13&6.06&3.87&3.96&4.36&4.42\\

\hline\hline
\end{tabular}
}
\end{table*}

\begin{table*}[htp]\centering
\caption{Statistical and systematic correlation matrices and relative uncertainties for the measured partial decay rates of  $D_s^+\to\eta^\prime \mu^+\nu_\mu$ in different
$q^2$ intervals.}
\label{tab:covb}
\begin{tabular}{ccccccc}\hline\hline
\multicolumn{7}{c}{Statistical correlation matrix}\\
\multirow{2}{*}{$\rho_{ij}^{\rm stat}$}&\multicolumn{3}{c}{$D_s^+\to\eta^\prime_{\eta\pi^+\pi^-}\mu^+\nu_\mu$}&\multicolumn{3}{c}{$D_s^+\to\eta^\prime_{\gamma\pi^+\pi^-}\mu^+\nu_\mu$}\\
&1&2&3&1&2&3\\
\hline
1	&1.000	&-0.086	&0.004	&0.000	&0.000	&0.000	\\
2	&&1.000	&-0.088	&0.000	&0.000	&0.000	\\
3	&&&1.000	&0.000	&0.000	&0.000	\\
1	&&&&1.000	&-0.065	&-0.018	\\
2	&&&&&1.000	&-0.073	\\
3	&&&&&&1.000	\\
\hline
Stat (\%)&17.08&18.50&27.84&14.04&12.58&29.05\\

\hline
\multicolumn{7}{c}{Systematic correlation matrix}\\
\multirow{2}{*}{$\rho_{ij}^{\rm syst}$}&\multicolumn{3}{c}{$D_s^+\to\eta^\prime_{\eta\pi^+\pi^-}\mu^+\nu_\mu$}&\multicolumn{3}{c}{$D_s^+\to\eta^\prime_{\gamma\pi^+\pi^-}\mu^+\nu_\mu$}\\
&1&2&3&1&2&3\\
\hline
1	&1.000	&0.730	&0.615	&0.592	&0.519	&0.274	\\
2	&&1.000	&0.960	&0.500	&0.488	&0.502	\\
3	&&&1.000	&0.423	&0.464	&0.533	\\
1	&&&&1.000	&0.631	&0.460	\\
2	&&&&&1.000	&0.168	\\
3	&&&&&&1.000	\\
\hline
Syst (\%)&4.22&4.74&5.98&4.43&6.46&8.19\\

\hline\hline
\end{tabular}
\end{table*}

\end{document}